\documentclass[11pt]{article}
\usepackage{amsmath}
\usepackage{amsthm}
\usepackage{amssymb}
\usepackage{verbatim}
\usepackage{dsfont}
\usepackage{bm}
\usepackage[margin=1in]{geometry}
\usepackage{algorithm}
\usepackage{hyperref}
\usepackage{cleveref}
\usepackage{scalerel}

\usepackage{thm-restate}

\newcommand{\bdisp}{\begin{displaystyle}}
\newcommand{\edisp}{\end{displaystyle}}

\renewcommand{\Pr}{\operatorname*{\mathbb{P}}}

\newcommand{\poly}{\operatorname*{\mathrm{poly}}}

\renewcommand{\d}{\mathrm{d}}

\newcommand{\eps}{\epsilon}

\newcommand{\del}{\delta}
\newcounter{nTheorems}

\newtheorem{theorem}[nTheorems]{Theorem}

\newtheorem{lemma}[nTheorems]{Lemma}

\newtheorem{fact}[nTheorems]{Fact}

\theoremstyle{definition}
\newtheorem{definition}[nTheorems]{Definition}

\usepackage[utf8]{inputenc}

\begin{document}
\title{A Generalized Trace Reconstruction Problem:\\ Recovering a String of Probabilities}
\author{ Joey Rivkin\\Stanford \and Gregory Valiant \\Stanford \and Paul Valiant\\Purdue}

\date{}
\maketitle

\begin{abstract}
We introduce the following natural generalization of \emph{trace reconstruction}, parameterized by a deletion probability $\del \in (0,1)$ and length $n$: There is a length $n$ string of probabilities, $S=p_1,\ldots,p_n,$ and each ``trace'' is obtained by 1) sampling a length $n$ binary string whose $i$th coordinate is independently set to 1 with probability $p_i$ and 0 otherwise, and then 2) deleting each of the binary values independently with probability $\del$, and returning the corresponding binary string of length $\le n$.  The goal is to recover an estimate of $S$ from a set of independently drawn traces.  In the case that all $p_i \in \{0,1\}$  this is the standard trace reconstruction problem.  We show two complementary results. First, for worst-case strings $S$ and any deletion probability at least order $1/\sqrt{n}$, no algorithm can approximate $S$ to constant $\ell_\infty$ distance or $\ell_1$ distance $o(\sqrt n)$ using fewer than $2^{\Omega(\sqrt{n})}$ traces.  Second---as in the case for standard trace reconstruction---reconstruction is easy for \emph{random} $S$: for any sufficiently small constant deletion probability, and any $\epsilon>0$, drawing each $p_i$ independently from the uniform distribution over $[0,1]$, with high probability $S$ can be recovered to $\ell_1$ error $\epsilon$ using $\poly(n,1/\epsilon)$ traces and computation time.  
We show indistinguishability in our lower bound by regarding a complicated alternating sum (comparing two distributions) as the Fourier transformation of some function evaluated at $\pm \pi,$ and then showing that the Fourier transform decays rapidly away from zero by analyzing its moment generating function.

\end{abstract}

\section{Introduction}
\emph{Trace reconstruction} is the problem of recovering a length $n$ binary string, $T$, from a set of independent \emph{traces}, where each trace is generated from $T$ by independently deleting each bit with probability $\del \in (0,1)$ and then returning the concatenation of the bits that were not deleted.  Since the introduction of this problem by Batu, Kannan, Khanna, and McGregor~\cite{Batu2004reconstructing} twenty years ago, it has received significant attention and yet remains surprisingly open.  The  best known upper bounds show that the problem can be solved using $\exp(\tilde{O}(n^{1/5}))$ traces, and the best known lower bounds show that $\tilde{\Omega}(n^{3/2})$ traces are necessary---both results due to Chase~\cite{chase2021new,chase2021separating}.  Indeed, beyond our inability to rigorously shrink this gap between upper and lower bounds, we seem to currently lack intuition for what the right answer should be---whether recovery should require polynomial, or super-polynomially many traces.  This is despite the fact that we know a near-optimal (though computationally expensive) algorithm: return the string that maximizes the likelihood of the traces~\cite{cheng2024kmerbased}.  

In this paper, we introduce a natural generalization of the trace reconstruction problem that relaxes the requirement that the true underlying sequence is binary. While the initial goal of investigating this new model was to provide some insights into the standard trace reconstruction problem, we believe that it is a well motivated and interesting problem in its own right.  

\begin{definition}[Generalized Trace Reconstruction]
    The \emph{generalized trace reconstruction} problem is defined in terms of a \emph{deletion probability} $\del \in (0,1)$.  Given a length $n$ string of probabilities, $S=p_1,\ldots,p_n$ with $p_i \in [0,1]$, each \emph{trace} is generated as follows:
    \begin{enumerate}
    \item  Generate a length $n$ binary string, $T=t_1,\ldots,t_n$, by independently setting each $t_i=1$ with probability $p_i$ and 0 otherwise.
    \item Delete each bit of $T$ independently with probability $\del$, and return the concatenation of the bits that were not deleted, which will be a binary string of length $\le n$.
    The goal is to recover an approximation of $S$ from a set of independently generated traces.
    \end{enumerate}
\end{definition}

Beyond the theoretical appeal of this generalized problem, it also seems to accurately model some of the initial motivating settings of trace reconstruction.  For example, the problem of reconstructing some reference genome given  degraded/deleted sequences/traces is most naturally formulated where each location in the reference genome corresponds to a probability as opposed to having a discrete value. This probability captures both the possibility of location-specific mutations (which occur both across individuals of a population, as well as across cells within a given individual), as well as location/site specific measurement error.  \medskip

 Our first result shows that, for worst-case inputs, every algorithm that can learn the true string to small constant $\ell_\infty$ distance, or $\ell_1$ distance $o(\sqrt n)$  with high probability over the randomness of the traces, must use at least $e^{\Omega(\sqrt{n})}$ traces.  This result holds even for deletion probabilities as small as $\Omega(1/\sqrt{n})$:

\begin{theorem}\label{thm:lower_bound}
There exist a pair of length $n$ sequences $S=p_1,\ldots,p_n$ and $S'=p'_1,\ldots,p'_n$ with constant $\ell_\infty$ distance and $\ell_1$ distance $\Theta(\sqrt{n})$, and an absolute  constant $c$ such that for any deletion probability $\del \geq \frac{c}{\sqrt{n}}$---and in particular, for all constant deletion probabilities---the distribution of traces drawn from $S$ versus $S'$ have total variation distance $e^{-\Omega(\sqrt{n})}$.
\end{theorem}

Complementing this strong negative result, we show that generalized trace reconstruction is easy \emph{on average}, in analogy to the standard trace reconstruction problem:

\begin{theorem}\label{thm:upper_bound}
Let $S=p_1,\ldots,p_n$ be chosen by drawing each $p_i$ independently from the uniform distribution over $[0,1]$.
For any constant deletion probability $\del \le 10^{-7},$ and desired accuracy $\epsilon>0$, there exists an algorithm for the generalized trace reconstruction problem which recovers $S$ to $\ell_1$ distance at most $\epsilon,$ using $\poly(n,1/\epsilon)$ traces and computation, and succeeds with probability at least $1-1/\poly(n)$ over the randomness of $S$ and the traces.
\end{theorem}

While we make no effort to optimize the degree of the polynomial bound on the number of traces and runtime required in the random setting, we note that a polynomial dependence is necessary.  Any algorithm that recovers $S$ to $\ell_1$ distance $\eps$ must learn a majority of the coordinates of $S$ to error $O(\eps/n),$ which requires at least order $n^2/\epsilon^2$ traces even in the case that the deletion probability $\delta = 0.$

\subsection{Discussion}
One of the core challenge in understanding the difficulty of the standard trace reconstruction problem is the discreteness---both the combinatorial nature of the deletions, and the constraint that each index of the original string is either 0 or 1.  Our formulation of the generalized setting preserves the combinatorial structure of deletions, and simply relaxes the binary nature of the underlying string.  From this vantage point, our super-polynomial lower bound for worst-case reconstruction may be evidence that standard trace reconstruction also requires many traces.  

Previous super-polynomial lower bounds for trace reconstruction and related problems either apply to restricted classes of algorithm (e.g. returning a function of the average trace~\cite{holenstein2008trace, de2017optimal} or  generalizations of this~\cite{cheng2024kmerbased}), or  apply to variants with additional structure confounding the deletions, such as  the ``population recovery'' variant of Ban, Chen, Freilich, Servedio, and Sinha where each trace is sampled from a distribution over strings, and then deletions are applied~\cite{ban2019beyond}.  In this sense, our lower bound seems to lie closest to the standard trace reconstruction problem.  Indeed, it is tempting to explore the natural interpolations between standard trace reconstruction and our generalization: if all the probabilities, $p_i$, lie in a small discrete set, or are multiples of $1/k$ for some parameter $k$, do strong lower bounds still apply? 
 What about if all but $k$ elements $p_i$ are required to be 0 or 1? 

Our positive results in the randomized setting also may hint that relaxing the binary nature of the true string might not make the problem that much more difficult.  That said, our efficient recovery algorithm in the randomized setting is significantly different than the recovery techniques that have appeared previously in the literature, which seem to crucially leverage the discreteness of the elements of $S$.  Indeed, the approaches to average-case recovery typically recover $S$ iteratively, leveraging the knowledge of $p_1,\ldots,p_i$ to identify specific regions of the traces, and ultimately recover $p_{i+1}$.  It seems hard to naively apply these techniques to our setting without the recovery error of subsequent $p_i$'s compounding geometrically.

\subsection{Related Works}
The problem of trace reconstruction was introduced in 2004~\cite{Batu2004reconstructing}, and built on several earlier papers on closely related problems of recovering strings from their subsequences~\cite{levenshtein2001efficient, levenshtein2001efficient2}.  Since then, it has enjoyed significant interest from the TCS, probability theory, and combinatorics communities.  Despite this study, we still do not understand the computational or information theoretic properties of the problem.  

On the side of upper bounds for constant deletion probability, the 2021 result of Chase \cite{chase2021separating} showed the current state-of-the-art (worst case) upper bound of $\exp{(\Tilde{O}(n^{1/5}))}$. 
 This improved upon the previous upper bounds,  an $\exp{(\Tilde{O}(\sqrt{n}))}$ trace algorithm in 2008~\cite{holenstein2008trace} and $\exp{(\Tilde{O}(n^{1/3}))}$ in 2017 \cite{nazarov2017trace,de2017optimal}.  On the lower bound side, the relatively recent papers of Holden and Lyons~\cite{holden2020lower} and Chase~\cite{chase2021new} showed first a $\Tilde{\Omega}(n^{5/4})$ and then $\Tilde{\Omega}(n^{3/2})$ lower bound. 

 Motivated by the extreme gap between the upper and lower bounds, there has also been significant work proving strong lower bounds again various natural restricted classes of algorithm.  These include an $\exp(n^{1/3})$ lower bound against ``mean-based'' algorithms, that return a function of the \emph{average} trace~\cite{holenstein2008trace,de2017optimal}, and very recent lower bound of $\exp(n^{1/5})$ for ``k-mer'' based algorithms that generalize mean-based algorithms~\cite{cheng2024kmerbased}. Both of these results showed that the best-known upper bounds were essentially optimal for the type of algorithms they analyze. 
 In the latter case, this shows that the $\exp(n^{1/5})$ result cannot be improved without considering significantly different algorithms.

Beyond the worst-case setting, trace reconstruction has also been considered in the \emph{average} case---where the true string is generated uniformly at random---beginning with the earliest papers on trace reconstruction ~\cite{Batu2004reconstructing, kannan2005more, viswanathan2008improved, holden2018subpolynomial, peres2017average, rubinstein2022average}.
In this setting, trace reconstruction is easy, requiring a \emph{subpolynomial} number of traces, with high probability.  Many earlier papers hinted at a strong connection between the worst-case sample complexity and the average-case sample complexity; this was formalized in the 2022 paper of Rubinstein~\cite{rubinstein2022average} who showed that an algorithm that uses $\exp(f(n))$ traces and succeeds in the worst case (even for ``shifted'' traces), yields an algorithm for the average case that requires only $\exp(\Theta(f(\log n)))$ traces.

There have also been significant recent efforts to explore natural variants of the trace reconstruction problem---in many cases motivated by the goal of better understanding why it has been so difficult to make progress on trace reconstruction.  This includes the work of \cite{chen2020polynomial} which showed trace reconstruction is easy in a ``smoothed'' sense.  Namely, in the setting where a worst-case input is chosen but then undergoes random perturbations and the goal is to recover the perturbed string \cite{chen2020polynomial}.  The \emph{approximate} trace reconstruction problem was introduced by \cite{davies2021approximate}, and subsequent work demonstrated that a constant number of traces suffice to reconstruct random inputs up to a small edit distance \cite{chase2021approximate, chen2022near, chakraborty2021approximate}. Many other variants have been proposed and studied including matrix reconstruction \cite{krishnamurthy2021trace}, circular trace reconstruction~\cite{narayanan2020circular}, and coded trace reconstruction \cite{cheraghchi2020coded, brakensiek2020coded}.

Most similar to our work is the FOCS'19 paper of Ban, Chen, Freilich, Servedio, and Sinha on ``population recovery''~\cite{ban2019beyond}.  They considered the problem of learning a distribution over length $n$ strings, given a set of traces that have been drawn by first sampling a string according to the distribution, and then drawing a trace from that string.  They showed that  distributions supported on at most $o(\log n/ \log \log n)$ strings can be learned with roughly $\exp(\sqrt{n})$ traces, and that there are distributions supported on $\ell < \sqrt{n}$ strings that provably require $\exp(\ell)$ traces to learn to nontrivial accuracy.   Our lower bound can be viewed within this framework as the problem of extending such strong lower bounds to the restrictive setting where the distribution in question is restricted to correspond to  flipping a sequence of $n$ coins, each of whose probabilities have been fixed.  Finally, we note that the lower bound construction in~\cite{ban2019beyond} is superficially similar to ours---with both involving the binomial distribution.  That said, the lower bound distributions of~\cite{ban2019beyond} are supported on strings containing at most a single 1, and hence the analysis corresponds to showing the indistinguishability of two integer-valued distributions (representing the location of the nonzero entry). In our case, by contrast, we can expect to see not just a single 1, but nearly $\sqrt{n}$ nonzero locations (with probability $e^{-O(\sqrt{n})}$); and these many locations interact with the deletion channel in intricate ways, requiring a new probabilistic analysis to get our $e^{\Omega(\sqrt{n})}$ trace lower bound.

\section{Techniques}

\subsection{Lower Bound}\label{sec:lower-techniques}
 Our lower bound constructs two length $n+1$ strings of probabilities---$S_e$ which is only nonzero on the \emph{even} locations in the string, and $S_o$ which is nonzero only on the \emph{odd} locations.  We describe the construction which is scaled by a parameter $\alpha$ which can be as large as $\Theta(\sqrt{n})$:
 \begin{equation*}
S_e(i) = 
\begin{cases} 
\alpha\,bin(n,\frac{1}{2},i)=\alpha\,{n\choose i} p^i (1-p)^{n-i}, & \text{if } i \text{ is even} \\ 
0, & \text{if } i \text{ is odd}
\end{cases}
\end{equation*}
 \begin{equation*}
S_o(i) = 
\begin{cases} 
0, & \text{if } i \text{ is even} \\ 
\alpha\,bin(n,\frac{1}{2},i)=\alpha\,{n\choose i} p^i (1-p)^{n-i}, & \text{if } i \text{ is odd}
\end{cases}
\end{equation*}
To show that the deletion channel transforms $S_e$ and $S_o$ into distributions of indistinguishable traces, we analyze the following ``partial deletion'' process: choose a (binomially distributed) random number $r\leftarrow Bin(\frac{n}{2},\frac{1}{2})-\frac{n}{4}$ between $\pm\frac{1}{4}n$ and, given a sample $x$ from either $S_e$ or $S_o$, return bits $r+\{\frac{1}{4}n+1,\ldots,\frac{3}{4}n\}$ of $x$, plus, separately, the initial $(\frac{1}{4}n+r)$ bits of $x$ \emph{after} they have gone through the deletion channel, and the final $(\frac{1}{4}n-r)$ bits of $x$ after they have gone through the deletion channel.  (This ``partial deletion'' process only makes the distinguishing problem easier, as one could always apply the deletion process to the middle bits and concatenate the first, middle, and last bits to generate an instance from the actual deletion channel.)  This partial deletion process corresponds to \emph{translating} the realization of the flips of the middle coins by a randomly sampled binomial; this allows us to apply the tools of convolution and Fourier transforms for the analysis.  Ultimately, we will show that under this partial deletion channel, the difference in distributions of traces from $S_o$ and $S_e$ is bounded by $e^{-\Omega(\sqrt{n})}$.

Each term in the expression for the statistical distance between these partial deletion channel traces from $S_e$ versus $S_o$ can be expressed as an alternating sum of an expression involving products of Binomial coefficients (see Equation~\ref{eq:r-func-lemma}), and our goal is to show all these terms are very small. As a motivating example, consider the alternating sum $\sum_{r=0}^n (-1)^r bin(n,\frac{1}{2},r)$, where $bin(n,\frac{1}{2},r)={n\choose r}2^{-n}$ is the binomial probability. This sum is famously always 0; but what if we modify it by taking $k^{\textrm{th}}$ powers of each term, as in \[\sum_{r=0}^n (-1)^r bin(n,\frac{1}{2},r)^k\] Can we show that, even with a higher exponent, $k>1$, this alternating sum still almost exactly cancels out? More generally, for a sequence of offsets $\ell_1,\ldots,\ell_k$, can we show that the following alternating sum of the $k$-way product of binomials almost exactly cancels: \[\sum_{r=0}^n (-1)^r \prod_{j=1}^k bin(n,\frac{1}{2},r+\ell_j)^k\]

Alternating sums of products of binomial coefficients can be expressed as (generalized) hypergeometric functions, hinting at plentiful structure; but for degree greater than 2 or 3, simplifications quickly become intractable. 

Instead, we view the alternating sum of a function, $\sum_r (-1)^r y(r)$ as being the frequency $\pm \pi$ evaluation of the \emph{Fourier transform} of $y$. And instead of evaluating the Fourier transform of $y$ at precisely the frequency $\pi$, we instead show that the Fourier transform is \emph{rapidly decaying away from $0$}, and therefore small at frequency $\pi$. To bound the decay of the Fourier transform of $y$, we estimate the \emph{moment generating function} of the Fourier transform to bound its tails. The Fourier transform converts the \emph{product} of several binomial coefficients into a corresponding \emph{convolution}; but the moment generating function transforms this convolution back into a product, which allows us to analyze how convergence of alternating sums of products \emph{gracefully} deteriorates as we multiply more terms. See Lemma~\ref{lem:fourier-bound} for the full analysis, leading to the proof of our lower bound, Theorem~\ref{thm:lower_bound}, expressed more specifically in Section~\ref{sec:lower} as Theorem~\ref{thm:lower-detail}.  We are not aware of similar analyses in the literature, and believe this sort of use of the moment generating function in the Fourier domain might be of broader use.

\subsection{Algorithms and Analysis}

Metaphorically, the trace reconstruction problem can be seen as analogous to the prominent robotics problem known as ``SLAM'': simultaneous localization and mapping. A robot wants to draw a map of its surroundings based on what it sees around it; but in order to fill in the right part of a map, it needs to know where it is; but in order to know where it is based on what it sees, it needs a map.... The inherently self-referential nature of this problem prompts the name \emph{simultaneous} localization and mapping. Analogously, in the trace reconstruction problem, given a portion of a trace, if we knew where in the original string it came from, then we could improve our ``map'' of the original string; by contrast, if we had a good map of the original string, we could use this map to easily compute the right alignment. This ``chicken and egg" problem prompts the flavor of many of the prior algorithms for the average-case trace reconstruction problem  (and many prominent SLAM algorithms): incrementally taking small steps along the input data, alternating between using the previously estimated ``map" to estimating our current location, and then using this location estimate to update our estimate of the ``map".

In the average-case trace reconstruction setting, prior approaches keep the error in the process  small enough to accurately snap estimates to either 0 or 1, and thereby prevent the accumulation of errors.  This is not possible in our present setting, where we are estimating the sequence of probabilities $p_1,\ldots,p_n$. Indeed, what we will strive for in our algorithm is that the bits that we use for ``mapping'' are completely disjoint from the bits used for ``localization'', so that there is no possibility that one part of the algorithm introduces bias into the other. Unlike prior approaches that move left to right in the traces, repeatedly alternating between localization and mapping phases: to avoid any complicated sources of bias we have only a single round of each phase.

We provide a high level summary of our approach here. See Section~\ref{sec:alg} for full details of our algorithm and analysis. (Here for simplicity we assume the parameter $m$ of Section~\ref{sec:alg} equals $n$.)

We pick a chunk size $w=10000\log n$. We collect $n^2$ traces in a set $X$; and a much larger set of $n^{25}$ traces in a set $Y$ that we will use to ``localize" chunks of traces in $X$. For each trace in $X$, we will pick three consecutive size-$w$ chunks, called $L,M,R$ (for Left, Middle, and Right); we will \emph{not} look at $M$ at all, but will instead output $M$ only if $L$ and $R$ satisfy certain properties with respect to the huge set of traces in $Y$. (Recall that our goal is to not use ``mapping" chunks $M$ for any localization tasks, to avoid biasing our estimates of the probabilities $p_i$.)

Intuitively, we wish to output strings $M$ that contain \emph{no} deletions; such strings are easily assembled into a whole. We design an algorithm that outputs strings $M$ with no deletions with a small but non-negligible probability; and outputs strings $M$ that have deletions only exponentially rarely. Intuitively, the chunk $M$ is likely to have no deletions if the chunks $L,R$---that bracket it---match up with \emph{many} pairs of chunks in $Y$ at the same separation $w$, or smaller separations $<w$, but \emph{never} at a larger separation $>w$. See Algorithm 1 for details. Algorithm 2 takes the deletion-free chunks output by Algorithm 1 and assembles them into an essentially unbiased estimate of the probabilities $p_1,\ldots,p_n$.

As a brief overview of the analysis: Even if two chunks both have no deletions and come from the same chunk of the source string, $p_{i+1},\ldots,p_{i+w}$, the observed traces will likely be different, because each bit results from a coin flip of probability $p_{i+1},\ldots,p_{i+w}$ respectively; thus instead of requiring an exact match, we match up chunks by thresholding their Hamming distance at $5w/12$. Lemma~\ref{lem:hoeffding} analyzes concentration via the Hoeffding bound; and Lemma~\ref{lemma:match} takes a union bound over Lemma~\ref{lem:hoeffding} in the context of Algorithm 1. Lemma~\ref{lemma:lem6} shows that with high probability there will be many triples of consecutive chunks $L,M,R$ in the traces in $X$ that have no deletions, and Lemma~\ref{lemma:lem7} shows that, in this case, the chunk $M$ will be correctly recognized by Algorithm 1, except with inverse exponential probability. Lemmas~\ref{lem:8} and~\ref{lemma:onlyDF} characterize the types of deletions that might occur in $L,M,R$ and show that with high probability, Algorithm 1 will output only deletion-free chunks $M$. These pieces are then easily assembled into a proof of Theorem~\ref{thm:upper_bound}.

\vspace{-3mm}

\section{Hardness for Worst-Case Strings}\label{sec:lower}

The lower bound proof centers around an argument bounding an alternating sum via estimates of the moment generating function of its Fourier transform---see Lemma~\ref{lem:fourier-bound} and its proof. We use this result to bound the discrepancy between distributions of traces associated with strings $S_e,S_o$.  As described in Section~\ref{sec:lower-techniques}, $S_e$ and $S_o$, are length $n+1$ strings where $S_e$ is nonzero only on even indices, and $S_o$ is nonzero only on odd indices. The strings are  parameterized by a scaling parameter $\alpha$ that controls the distance between $S_e,S_o$, where $\alpha=\Theta(1)$ induces $S_e,S_o$ with constant $L_1$ distance; while $\alpha=\Theta(\sqrt{n})$ induces the much stronger setting for our lower bound where $S_e,S_o$ have constant $L_\infty$ distance. 

\begin{definition}\label{def:construction}
Define two sequences of $n+1$ probabilities, $S_e$ and $S_o$ indexed by $i\in\{0,\ldots,n\}$, where $S_e$ is nonzero on even indices, and $S_o$ is nonzero on odd indices. Letting $bin(n,p,k):={n\choose k} p^k (1-p)^{n-k}$ be the probability that $n$ flips of a $p$-biased coin results in $k$ heads, we define $S_e$ for even indices $i$ to equal $\alpha\,bin(n,\frac{1}{2},i)$ and 0 otherwise; and define $S_o$ for odd indices $i$ to equal $\alpha\,bin(n,\frac{1}{2},i)$ and 0 otherwise. 
\end{definition}

For the sake of symmetry, we consider $n$ odd, so that $S_e$ and $S_o$ differ by a reflection about the center. 
Even when $\alpha$ is as low as $\Theta(1)$, the sequences $S_e,S_o$ have constant $L_1$ distance, and thus can be distinguished via a constant number of samples. However, we show that when sent through a deletion channel with even a deletion probability $\del$ as small as $\Omega(\frac{1}{\sqrt{n}})$, the corresponding distributions of traces become essentially indistinguishable.

\begin{theorem}\label{thm:lower-detail}
There are constants $c,c'$ such that for any deletion probability $\del\geq \frac{c}{\sqrt{n}}$---and in particular, for all constant deletion probabilities---and for any scaling factor $\alpha\leq c'\sqrt{n}$ (in the construction of $S_e,S_o$ in Definition~\ref{def:construction}), the statistical distance between a trace from $S_e$ versus $S_o$ is $e^{-\Omega(\sqrt{n})}$.
\end{theorem}
As a trivial corollary, there is a constant $c''$ such that, for sufficiently large $n$, and any number of traces $t$, there is no algorithm that, given $t$ traces all from $S_e$ or all from $S_o$, can distinguish these two cases with probability better than $\frac{1}{2}+t \cdot e^{-c'' \sqrt{n}}$.

As described in Section~\ref{sec:lower-techniques}, to prove the theorem, we bound the statistical distance by first relating the deletion channel to an easier-to-analyze process that instead shifts a portion of the string by a random offset, yielding the following lemma.

\begin{lemma}\label{lem:r-func}
For deletion probability $\del$, the statistical distance between a trace from $S_e$ versus a trace from $S_o$ is at most the sum over all $k\in\{0,\ldots,\frac{n}{2}\}$ of the sum over all $k$-tuples of locations $\ell_1<\ell_2<\ldots<\ell_k$ in $\{1,\ldots,\frac{n}{2}\}$ that have identical parity, of the sum over all $z_-,z_+\in\{0,\ldots,\frac{n}{2}\}$ of the following expression
\begin{equation}\label{eq:r-func-lemma}\left|\sum_{r} (-1)^r\cdot bin( n,\frac{1}{2},r+\frac{1}{2}n)\cdot bin(\frac{1}{4}n+r,1-\del,z_-)\cdot bin(\frac{1}{4}n-r,1-\del,z_+)\prod_{j=1}^k \frac{\alpha\,bin(n,\frac{1}{2},r+\frac{1}{4}n+\ell_j)}{1-\alpha\,bin(n,\frac{1}{2},r+\frac{1}{4}n+\ell_j)} \right|,\end{equation}
up to a $O(1)$ multiplicative term and a $e^{-\Omega(n)}$ additive term.
\end{lemma}
\begin{proof}
An upper bound on the statistical distance is given by the statistical distance of the following ``partial deletion" process: choose a (binomially distributed) random number $r\leftarrow Bin(\frac{n}{2},\frac{1}{2})-\frac{n}{4}$ between $\pm\frac{1}{4}n$ and, given a sample $x$ from either $S_e$ or $S_o$, return bits $r+\{\frac{1}{4}n+1,\ldots,\frac{3}{4}n\}$ of $x$, plus, separately, the initial $(\frac{1}{4}n+r)$ bits of $x$ \emph{after} they have gone through a deletion channel, and the final $(\frac{1}{4}n-r)$ bits of $x$ after they have gone through a deletion channel.

The reason this process gives an upper bound on the statistical distance is that, given the output of this process, we can simulate the authentic deletion channel by deleting each bit from the middle segment with probability $\del$, and appending the initial and final strings of bits; by the information processing inequality, the statistical distance between two processes cannot increase if we run the data through the same transformation.

We thus analyze the statistical distance between $S_e$ and $S_o$ under this new process.

We immediately observe that $r$ will be within $\pm \frac{1}{8}n$ except with $e^{-\Omega(n)}$ probability; and given this, the initial $(\frac{1}{4}n+r)$ and final $(\frac{1}{4}n-r)$ bits each have probability of being nonzero bounded by the left tails of the binomial distributions $bin(n,\frac{1}{2},\leq \frac{n}{2}-\frac{n}{8})\leq bin(n,\frac{1}{2},\leq\frac{3}{8}n)=e^{-\Omega(n)}$. Thus the statistical distance is changed by at most $e^{-\Omega(n)}$ if, instead of receiving deletion channel traces of the initial $(\frac{1}{4}n+r)$ and final $(\frac{1}{4}n-r)$ bits, we instead assume these are strings of entirely 0s, and receive just their lengths instead. 

Thus we consider the (equivalent to within distance $e^{-\Omega(n)}$) statistical process where we sample $r\leftarrow Bin(\frac{n}{2},\frac{1}{2})-\frac{n}{4}$, and then return the $\frac{n}{2}$ bits $r+\{\frac{1}{4}n+1,\ldots,\frac{3}{4}n\}$ of a sample from either $S_e$ or $S_o$, and also receive two integers $z_-\leftarrow Bin(\frac{1}{4}n+r,1-\del)$ and $z_+\leftarrow Bin(\frac{1}{4}n-r,1-\del)$. Keep in mind $r$ is crucially \emph{not} returned in this process; instead, $z_-$ and $z_+$ are returned as fuzzy proxies for $r$, hiding the true offset if the deletion probability $\del$ is high enough.

Let $y$ be the length $\frac{1}{2}n$ string returned from the ``middle'' of the sample. Let $k$ denote the number of nonzero bits in $y$; we record their locations in $y$ as $\ell_1,\ldots,\ell_k\in\{1,\ldots,\frac{n}{2}\}$.

Since bit $\ell_j$ in $y$ has location $r+\frac{1}{4}n+\ell_i$ in the original string, the probability of this bit being 1 equals $bin(n,\frac{1}{2},r+\frac{1}{4}n+\ell_j)$. Thus, fixing $r,z_-,z_+,k,$ and $\{\ell_1,\ldots,\ell_k\}$, we can compute the probability of this outcome arising from the $S_e$ (respectively $S_o$) process: if the parity of all $r+\ell_j$ is even (respectively odd), we thus compute the probability of $r,z_-,z_+$ being drawn, and then of the nonzero bits in the overall string being exactly bits $\ell_1,\ldots,\ell_k$ from the middle segment as being
\[ bin( n,\frac{1}{2},r+\frac{1}{2}n)\cdot bin(\frac{1}{4}n+r,1-\del,z_-)\cdot bin(\frac{1}{4}n-r,1-\del,z_+)\prod_{j=1}^k \frac{\alpha\,bin(n,\frac{1}{2},r+\frac{1}{4}n+\ell_j)}{1-\alpha\,bin(n,\frac{1}{2},r+\frac{1}{4}n+\ell_j)} \prod_{\substack{j=0\\j\textrm{ even}}}^n (1-\alpha\,bin(n,\frac{1}{2},j))\]where we change ``$j$ even" in the range of the final product to ``$j$ odd" for $S_o$ instead of $S_e$.

We observe that, for odd $n$ the final term $\prod_{\substack{j=0\\j\textrm{ even}}}^n (1-\alpha\,bin(n,\frac{1}{2},j))$ has identical value in both the even $j$ and the odd $j$ case by symmetry; and in both cases the product is $O(1)$, so up to constant factors, we can drop this term, which we do.

Thus, the difference between probabilities of observing $z_-,z_+,\{\ell_1,\ldots,\ell_k\}$ under the $S_e$ versus $S_o$ cases is (up to sign, and up to the additive $e^{-\Omega(n)}$ term from earlier) exactly the alternating sum over $r$ of $(-1)^r$ times the above equation, as claimed in the lemma.
\end{proof}

Our overall strategy to bound the terms of Equation~\ref{eq:r-func-lemma} will be to view the sum over $r$ of $(-1)^r$ times some function $y(r)$ instead as the Fourier transform of this function $y$, evaluated at angle $\xi=\pi$. Since the function of Equation~\ref{eq:r-func-lemma} is a \textit{product} of several terms, its Fourier transform is the \textit{convolution} of the Fourier transform of each term. And our strategy will be to show that these Fourier transforms decay away from 0, so that when they are convolved and evaluated at angle $\pi$, their contributions this far from 0 can all be bounded as exponentially small in $\sqrt{n}$. The below lemma is the main technical step, bounding the Fourier transform of the final $k$-way product from Equation~\ref{eq:r-func-lemma}.

We summarize the main steps in the proof of Lemma~\ref{lem:fourier-bound}, as each step involves a significant transformation. We start by using a power series for $\frac{z}{1-z}$ to reexpress the fractions in Equation~\ref{eq:r-func-eq} as an infinite sum of positive powers of the binomial function, in Equation~\ref{eq:power-series}. We then reexpress this $k$-way \textit{product} as a $k$-way \textit{convolution} in the Fourier domain, in Equation~\ref{eq:first-convolution}. However, since the Fourier transform of a function supported on the integers is defined modulo $2\pi$, we instead relate this Fourier transform $f(\xi)$---where $\xi$ is a real number mod $2\pi$---to a function over all the reals, $h(\xi')$, essentially ``unwrapping'' $f$ to produce $h$, and taking absolute values of Fourier transforms to leave us with a real-valued function: see Equation~\ref{eq:h}. We then work to compute the moment generating function of $h$. (The moment generating function would make no sense mod $2\pi$, which is part of the reason we had to change the domain to the entire real line.) We first compute the moment generating function of the Fourier transform just of $bin(n,\frac{1}{2},r)$ in Equation~\ref{eq:mgf-binomial}, and then we bound this moment generating function in Equation~\ref{eq:g-bound}, before using this bound to get a bound on the moment generating function of the full $h$, in Equation~\ref{eq:full-mgf}. We use this moment generating function to bound the tails of the overall Fourier transform, in Equation~\ref{eq:mgf-bound}. Finally, we show that the magnitude of our Fourier transform $h$ is decreasing away from 0, and use this to convert the \emph{cumulative} bound on the tails of the Fourier transform into the desired \emph{pointwise} bound on the Fourier transform, yielding the lemma.

\begin{lemma}\label{lem:fourier-bound}
Given $k$ distinct locations $\ell_1,\ldots,\ell_k\in\{1,\ldots,\frac{n}{2}\}$, define the Fourier transform with respect to $r$, defined on angles $\xi\in[-\pi,\pi]$, of the product function from Equation~\ref{eq:r-func-lemma}, where we interpret the binomial pdf to be 0 if its third argument is out of range:
\begin{equation}\label{eq:r-func-eq}f(\xi):=\sum_r e^{i\xi r}\prod_{j=1}^k \frac{\alpha\,bin(n,\frac{1}{2},r+\frac{1}{4}n+\ell_j)}{1-\alpha\,bin(n,\frac{1}{2},r+\frac{1}{4}n+\ell_j)}\end{equation}

We claim that, if $|\xi|\geq 2$ and $\alpha\leq \frac{\sqrt{n}}{4 e^2\sqrt{2\pi}}$, then $|f(\xi)|\leq 2\cdot e^{-\sqrt{n}}$.
\end{lemma}

\begin{proof}
We first point out that we can simplify the fraction $\frac{\alpha\,bin(n,\frac{1}{2},r+\frac{1}{4}n+\ell_j)}{1-\alpha\,bin(n,\frac{1}{2},r+\frac{1}{4}n+\ell_j)}$. If, for the moment, we let $z=\alpha\,bin(n,\frac{1}{2},r+\frac{1}{4}n+\ell_j)$, where $z\in[0,1)$, then $\frac{z}{1-z}=\frac{1}{1-z}-1=\sum_{b=1}^\infty z^b$, which will let us remove fractions from the expression being Fourier transformed. As a minor technical issue, we will also replace $\frac{1}{4}n$ with 0 as it is added to $r$ in the binomial expression; this has the effect of shifting $r$ by a constant in the function being Fourier transformed, which will not affect the magnitude of the resulting Fourier transform, only its phase. Thus

\begin{equation}\label{eq:power-series}[|f(\xi)|=\left|\sum_r e^{i\xi r}\prod_{j=1}^k \sum_{b=1}^\infty (\alpha\,bin(n,\frac{1}{2},r+\ell_j))^b\right|\end{equation}

We use $\mathcal{F}$ to denote the Fourier transform, always over the variable $r$, and where we use $\mathcal{F}_\xi$ to emphasize that the output of the Fourier transform will be expressed in terms of a (new) variable $\xi$. Letting $\circledast$ denote convolution modulo $2\pi$ (since the Fourier transform of a function supported on the integers is defined modulo $2\pi$), we have, since the Fourier transform of a product is the convolution of the Fourier transforms of each term: \begin{equation}\label{eq:first-convolution}|f(\xi)|=\left|\mathop{\scaleobj{1.5}{\circledast}}_{j=1}^k \mathcal{F}_\xi\Big(\sum_{b=1}^\infty (\alpha\,bin(n,\frac{1}{2},r+\ell_j))^b\Big)\right|\end{equation}

We introduce a modified version of $f$, denoted $h(\xi')$ that is in terms of a real variable $\xi'$ (in contrast to $\xi$, which is interpreted mod $2\pi$); we will then bound $f$ in terms of $h$, and then bound $h$: let \begin{equation}\label{eq:h}h(\xi')=\left(\sum_{b=1}^\infty \left|\mathcal{F}_{\xi'}(\alpha\,bin(n,\frac{1}{2},r))\right|^{*b}\right)^{*k}\end{equation} 
where in Equation~\ref{eq:h} the Fourier transform is interpreted as returning a function supported within the interval $[-\pi,\pi]$, and the superscripts ${}^{*b}$ and ${}^{*k}$ denote $b$-way convolution and $k$-way convolution respectively, both over the reals (and, crucially, this convolution is \emph{not} modulo $2\pi$).

We first show that $h$ is an ``unwrapped" version of $f$, in that \[|f(\xi)|\leq \sum_{s=-\infty}^\infty h(\xi+2\pi s)\] To show this, we observe that \[|f(\xi)|\leq\mathop{\scaleobj{1.5}{\circledast}}_{j=1}^k\left| \mathcal{F}_\xi\Big(\sum_{b=1}^\infty (\alpha\,bin(n,\frac{1}{2},r+\ell_j))^b\Big)\right|=\mathop{\scaleobj{1.5}{\circledast}}_{j=1}^k\left| \mathcal{F}_\xi\Big(\sum_{b=1}^\infty (\alpha\,bin(n,\frac{1}{2},r))^b\Big)\right|\leq \mathop{\scaleobj{1.5}{\circledast}}_{j=1}^k\sum_{b=1}^\infty \left| \mathcal{F}_\xi(\alpha\,bin(n,\frac{1}{2},r)^b)\right|\]
where this last expression is bounded by $\left(\sum_{b=1}^\infty  \left|\mathcal{F}_\xi(\alpha\,bin(n,\frac{1}{2},r))\right|^{\circledast b}\right)^{\circledast k}$. Replacing the circular convolution operator $\circledast$ by (regular) convolution $*$ yields exactly the expression for $h$ of Equation~\ref{eq:h}, meaning that if we sum Equation~\ref{eq:h} over all $\xi'$ that are equal to a given $\xi$ mod $2\pi$, the resulting sum will bound $|f(\xi)|$, as claimed.

Before bounding the moment generating function of $h$, we point out that the Fourier transform defining $h$, namely $\mathcal{F}_{\xi'}(bin(n,\frac{1}{2},r)$ can be easily computed (where we will multiply by $\alpha$ later). The binomial function is the convolution of $n$ fair coin flips; and thus its Fourier transform is the ${n}^\textrm{th}$ power of a single coin flip, whose Fourier transform (up to phase, which does not matter) is $\cos(\frac{\xi'}{2})$, for $\xi'\in [-\pi,\pi]$.

The moment generating function of $\mathcal{F}_{\xi'}(bin(n,\frac{1}{2},r)$, for odd $n$, is thus \begin{equation}\label{eq:mgf-binomial}g(t):=\int_{-\pi}^{\pi} \cos(\frac{\xi'}{2})^{n}\,e^{t\xi'}\,d\xi'=\frac{2^{n+1} \cosh(\pi t)}{{n\choose \frac{n-1}{2}}\frac{n+1}{2}\prod_{j=0}^{(n-1)/2} (1+\frac{t^2}{(j+\frac{1}{2})^2})}=\frac{2^{n+1}}{{n\choose \frac{n-1}{2}}\frac{n+1}{2}}\prod_{j=\frac{n+1}{2}}^{\infty} (1+\frac{t^2}{(j+\frac{1}{2})^2})\end{equation}

We bound this product, using the fact that $1+x\leq e^x$, and using the fact that we can bound the sums of inverse squares starting at $\frac{n}{2}+1$ by the corresponding integral starting at $\frac{n}{2}$, as $\prod_{j=\frac{n+1}{2}}^{\infty} (1+\frac{t^2}{(j+\frac{1}{2})^2})\leq e^{\frac{2t^2}{n}}$. Since the remaining part of the expression is bounded as $\frac{2^{n+1}}{{n\choose \frac{n-1}{2}}\frac{n+1}{2}}\leq \sqrt{\frac{8\pi}{n}}$, we have \begin{equation}\label{eq:g-bound}g(t)\leq e^{\frac{2t^2}{n}}\sqrt{\frac{8\pi}{n}}\end{equation}

Further, $\mathcal{F}_{\xi'}(\alpha\,bin(n,\frac{1}{2},r-\frac{n}{2}))^{*b}$ is the $b$-way convolution of $\mathcal{F}(\alpha\,bin(n,\frac{1}{2},r-\frac{n}{2}))$. Since the moment generating function of a convolution equals the product of the moment generating functions, we conclude that the moment generating function of $\mathcal{F}_{\xi'}(\alpha\,bin(n,\frac{1}{2},r-\frac{n}{2}))^{*b}$ equals $(\alpha\, g(t))^b$.

Thus the moment generating function of $\sum_{b=1}^\infty \mathcal{F}_{\xi'}(\alpha\,bin(n,\frac{1}{2},r-\frac{n}{2}))^{*b}$ equals $\sum_{b=1}^\infty (\alpha\,g(t))^b$. Provided $\alpha g(t)<1$, this geometric series sums to exactly $\frac{1}{1-\alpha\,g(t)}-1$.

Thus using our above bound on $g(t)$ we have
\begin{equation}\label{eq:full-mgf}[\mathrm{mgf}_t(h)= \left(\frac{1}{1-\alpha\,g(t)}-1\right)^k \leq \left(\frac{1}{1-\alpha\,e^{\frac{2t^2}{n}}\sqrt{\frac{8\pi}{n}}}-1\right)^k\end{equation}
where this bound is valid as long as the denominator stays positive.

Given this bound on the moment generating function of the Fourier transform $h$, we then plug in $t=\pm\sqrt{n}$ which yields a moment generating function $\leq 1$ in both cases, provided $\alpha\leq \frac{\sqrt{n}}{4 e^2\sqrt{2\pi}}$.

Explicitly, this moment generating function bound states that \begin{equation}\label{eq:mgf-bound}\int_{-\infty}^\infty h(\xi') e^{\sqrt{n}\xi'}\,d\xi'\leq 1\end{equation}

For $\xi'\geq 1$, the coefficient $e^{\sqrt{n}\xi'}$ is at least $e^{\sqrt{n}}$, and thus we conclude that $\int_1^\infty h(\xi')\,d\xi'\leq e^{-\sqrt{n}}$. Symmetrically, using $t=-\sqrt{n}$ yields $\int_{-\infty}^{-1} h(\xi')\,d\xi'\leq e^{-\sqrt{n}}$.

We observe that $h$ is monotonically decreasing away from 0: from the definition of $h$ in Equation~\ref{eq:h} and the Fourier transform of the Binomial function we have that $h(\xi')=\left(\sum_{b=1}^\infty (\alpha\,\cos(\frac{\xi'}{2}))^{*b}_{[-\pi,\pi]}\right)^{*k}$; we could reexpress this as a sum of convolutions; the Pr\'{e}kopa-Leindler inequality says that convolutions of log-concave functions are log-concave; thus since $\cos(\frac{\xi'}{2})$ is log-concave in the domain $[-\pi,\pi]$, we have that $h$ is the sum of log-concave functions, each symmetric about 0. Thus $h$ is decreasing away from 0.

The fact that $h$ is monotonically decreasing away from 0 says that for any $\xi'\geq 2$ we have that $h(\xi')\leq \int_{\xi'-1}^{\xi'} h(u)\,du$, with a corresponding relation for $\xi'\leq-2$. Thus, for any $\xi\in [-\pi,\pi]$ but where $|\xi|\geq 2$ we have that $|f(\xi)|\leq\sum_{s=-\infty}^\infty h(\xi+2\pi s)\leq \int_{\xi'\in \mathbb{R}/ (-2,2)} h(\xi')\,d\xi'\leq 2\cdot e^{-\sqrt{n}}$, thus proving the desired result.
\end{proof}

Having bounded the Fourier transform of the complicated $k$-way product in Equation~\ref{eq:r-func-lemma}, we can get relatively straightforward bounds on the Fourier transform of the rest of Equation~\ref{eq:r-func-lemma} and use this---after some arithmetic---to derive an overall bound on Equation~\ref{eq:r-func-lemma}.

\begin{lemma}\label{lem:r-func-bound}
If $\alpha\leq \frac{\sqrt{n}}{4 e^2\sqrt{2\pi}}$ then Equation~\ref{eq:r-func-lemma} from Lemma~\ref{lem:r-func} is always at most \[(n+1)(2\pi-2)\frac{(2\pi)^2}{(1-\del)^2} \max\{\frac{e^{-\frac{\del}{20}z_-}}{1-\del},\frac{e^{-\frac{\del}{20}z_+}}{1-\del},e^{-\frac{n}{150}}\}+4\frac{(2\pi)^2}{(1-\del)^2} e^{-\sqrt{n}}.\]
\end{lemma}

\begin{proof}
We compute the magnitude of the Fourier transform (with respect to $r$) of one of the binomial terms, using the general fact that, for parameter $c$ with $|c|<1$ we have $\sum_{n=k}^\infty {n\choose k} c^{n-k}=(1-c)^{-(k+1)}$: \[\hspace{-1cm}\mathcal{F}_{\xi}(bin(\frac{1}{4}n+r,1-\del,z_{-}))=\sum_r e^{i\xi r}bin(\frac{1}{4}n+r,1-\del,z_{-})=\sum_r e^{i\xi r}{\frac{n}{4}+r\choose z_{-}} (1-\del)^{z_{-}} \del^{\frac{n}{4}+r-z_{-}}=\frac{e^{-i\xi(\frac{n}{4}-z_{-})}(1-\del)^{z_{-}}}{(1-\del \cdot e^{i\xi})^{z_{-}+1}}\]

The magnitude of this is clearly maximized when $\xi=0$ in which case it has magnitude $\frac{1}{1-\del}$. Further, when $|\xi|\geq \frac{1}{3}$, we can easily check that $\frac{1-\del}{|1-\del\cdot e^{i\xi}|}\leq 1-\frac{\del}{20}\leq e^{-\frac{\del}{20}}$, leading to a bound of $|\mathcal{F}_{\xi}(bin(\frac{1}{4}n+r,\del,z_{-}))|\leq \frac{e^{-\frac{\del}{20}z_-}}{1-\del}$ when $|\xi|\geq \frac{1}{3}$.

Analogous bounds hold for the $z_+$ binomial, replacing $z_-$ by $z_+$ in the result.

And for the first binomial, $bin(\frac{n}{2},\frac{1}{2},r+\frac{n}{2})$ we already know that, up to phase, its Fourier transform equals $\cos(\frac{\xi}{2})^\frac{n}{2}$. This is always bounded by 1; and if $|\xi|\geq \frac{1}{3}$ then $\cos(\frac{\xi}{2})^\frac{n}{2}\leq e^{-\frac{n}{150}}$.

Thus the Fourier transform of the product of the three binomials equals the convolution of the Fourier transforms of each binomial; and, expressing the 3-way convolution as a double integral over domain $2\pi\times 2\pi$, we have that, for $|\xi|\geq 1$, at least one of the three arguments into the functions being convolved must be at least $\frac{1}{3}$, thus leading to a bound of $\frac{(2\pi)^2}{(1-\del)^2} \max\{\frac{e^{-\frac{\del}{20}z_-}}{1-\del},\frac{e^{-\frac{\del}{20}z_+}}{1-\del},e^{-\frac{n}{150}}\}$. And for all $\xi$ the Fourier transform is bounded by $\frac{(2\pi)^2}{(1-\del)^2}$.

We point out that, in the $k$-way product of Equation~\ref{eq:r-func-lemma}, we have $\alpha\,bin(n,\frac{1}{2},r+\frac{1}{4}n+\ell_j)\leq \frac{1}{2}$ (since binomial probabilities are bounded by $\sqrt{\frac{2}{\pi n}}$ and $\alpha\leq \frac{\sqrt{n}}{4 e^2\sqrt{2\pi}}$), and thus every term in the $k$-way product is $\leq 1$. Thus the Fourier transform of the $k$-way product---which involves summing over the domain of size $n+1$---trivially has magnitude at most $n+1$.

We combine this with the above bounds on the Fourier transform of the first 3 terms of Equation~\ref{eq:r-func-lemma} and the result of Lemma~\ref{lem:fourier-bound}, to conclude that the overall alternating sum---equaling the overall Fourier transform evaluated at $\xi=\pm\pi$---is bounded by $(n+1)(2\pi-2)\frac{(2\pi)^2}{(1-\del)^2} \max\{\frac{e^{-\frac{\del}{20}z_-}}{1-\del},\frac{e^{-\frac{\del}{20}z_+}}{1-\del},e^{-\frac{n}{150}}\}+2\frac{(2\pi)^2}{(1-\del)^2} 2 e^{-\sqrt{n}}$ for $k>0$. For $k=0$ we simply use the bound on the Fourier transform of the first 3 terms, evaluated at $\xi=\pm\pi$, namely $\frac{(2\pi)^2}{(1-\del)^2} \max\{\frac{e^{-\frac{\del}{20}z_-}}{1-\del},\frac{e^{-\frac{\del}{20}z_+}}{1-\del},e^{-\frac{n}{150}}\}$, thus proving the desired bound in all cases.
\end{proof}

Before we prove the theorem, we first show how to bound the number of possible arrangements of nonzero bits from a sample from $S_e$ or $S_o$ with high probability. We bound this via the $L_{1/2}$ norm of the relevant distributions.

\begin{fact}\label{fact:sqrt}
We may bound the sum of the square roots of binomial probabilities: \[\sum_{i=0}^n \sqrt{2^{-n}{n\choose i}}\leq (2\pi n)^{1/4}\]    
\end{fact}

\begin{lemma}\label{lem:sqrt}
The distribution on $n+1$ bit strings induced by $S_e$ (without deletions) has sum of the square roots of its probabilities bounded by $e^{\sqrt{\alpha}(2\pi n)^{1/4}}$; by symmetry the same bound applies to $S_o$.
\end{lemma}
\begin{proof}
Recall that for even $i$, the $i^{\textrm{th}}$ entry of $S_e$ is $S_e(i)=\alpha\,{n\choose i}2^{-i}$. Thus we can bound the sum of the square roots of this probability and its complement as \[\sqrt{S_e(i)}+\sqrt{1-S_e(i)}\leq 1+\sqrt{S_e(i)}\leq e^{\sqrt{S_e(i)}}\leq e^{\sqrt{\alpha\,2^{-n}{n\choose i}}}\]
Thus, since each element $i$ of the given distribution is independent, the sum of the square roots of the probabilities of the given distribution equals the product of the above expression over all $i$. By Fact~\ref{fact:sqrt} this is bounded by \[\prod_{i=0}^n e^{\sqrt{\alpha\,2^{-n}{n\choose i}}}\leq e^{\sqrt{\alpha}(2\pi n)^{1/4}}\]
\end{proof}

\begin{lemma}\label{lem:rare}
For the distribution on $n+1$ bit strings induced by $S_e$ before any deletions have occurred, the number of realizations that have probabilities $\geq e^{-\sqrt{n}/2}$ is at most $e^{\sqrt{n}/2}$, and encompasses all but at most $e^{\sqrt{\alpha}(2\pi n)^{1/4}-\sqrt{n}/4}$ of the total probability mass. By symmetry, the same bound applies to $S_o$.
\end{lemma}
\begin{proof}
Letting $p$ denote the probability distribution under discussion, and let $j$ index over its domain elements. Lemma~\ref{lem:sqrt} says that $\sum_j \sqrt{p(j)}\leq e^{\sqrt{\alpha}(2\pi n)^{1/4}}$. We use this to bound the total probability mass of elements of probability below $e^{-\sqrt{n}/2}$ as follows:
\[\sum_{j:p(j)<e^{-\sqrt{n}/2}}p(j)\leq \sum_j p(j) \frac{e^{-\sqrt{n}/4}}{\sqrt{p(j)}}\leq e^{\sqrt{\alpha}(2\pi n)^{1/4}-\sqrt{n}/4}\]
The remaining part of the lemma is trivial: the number of domain elements $j$ that have probabilities $\geq e^{-\sqrt{n}/2}$ is at most $e^{\sqrt{n}/2}$ since otherwise the probabilities would sum to more than 1.
\end{proof}

We now assemble the pieces to complete the proof of the theorem.

\begin{proof}[Proof of Theorem~\ref{thm:lower-detail}]
We prove the theorem for deletion probability $\del\leq \frac{1}{2}$: a larger deletion probability can only decrease the statistical distance, by the information processing inequality, since we can simulate deleting more bits from the trace.

We use Lemma~\ref{lem:r-func}'s characterization of the statistical distance between a trace from $S_e$ versus $S_o$: up to a constant multiplicative factor and an exponentially small additive error, the statistical distance is the sum over $k$-tuples of locations $\ell_1,\ldots,\ell_k$ where $S_o$ or $S_e$ could be nonzero, possibly shifted by $r\leftarrow Bin(\frac{n}{2},\frac{1}{2})-\frac{n}{4}$ and summed over samples $z_-\leftarrow Bin(\frac{1}{4}n+r,1-\del)$ and $z_+\leftarrow Bin(\frac{1}{4}n-r,1-\del)$, of the expression in Equation~\ref{eq:r-func-lemma}. We have bounded Equation~\ref{eq:r-func-lemma} in Lemma~\ref{lem:r-func-bound}.

Before summing up the bounds of Lemma~\ref{lem:r-func-bound} over all possibilities, we first describe some basic bounds on the values of the variables, which will allows us to limit the space of possibilities we sum over.

The value $r\leftarrow Bin(\frac{n}{2},\frac{1}{2})-\frac{n}{4}$ has magnitude at most $\frac{n}{8}$ except with probability $e^{-\Omega(n)}$. Since we assume $\del\leq \frac{1}{2}$, we thus have that $z_-\leftarrow Bin(\frac{1}{4}n+r,1-\del)$ and $z_+\leftarrow Bin(\frac{1}{4}n-r,1-\del)$ will have values at least $\frac{n}{16}$ except with probability $e^{-\Omega(n)}$. Henceforth we assume this does not happen.

We invoke Lemma~\ref{lem:rare} to conclude that, since $\alpha\leq \sqrt{n}/100$ then, except with probability $e^{-\Omega(\sqrt{n})}$, a realization of $S_e$ or of $S_o$ will be one of the $\leq e^{\sqrt{n}/2}$ possibilities encompassed by Lemma~\ref{lem:rare}. We thus sum up the bounds of Lemma~\ref{lem:r-func-bound} over all these possibilities, shifted by any possible offset $r\in\{-n/8,\ldots,n/8\}$, and over all $z_-,z_+\geq \frac{n}{16}$. For $\delta\in [\frac{320}{\sqrt{n}},\frac{1}{2}]$, the bound of Lemma~\ref{lem:r-func-bound} is $O(n\cdot e^{-\sqrt{n}})$. Thus, even summing this bound over the $e^{\sqrt{n}/2}poly(n)$ possibilities just described, the total discrepancy is $e^{-\Omega(\sqrt{n})}$. This proves the theorem.
\end{proof}

\section{Efficient Reconstruction of Random Strings}\label{sec:alg}

\vspace{-1mm}
In this section, we prove Theorem~\ref{thm:upper_bound}.  We begin by describing our algorithm for reconstructing a string $S=p_1,\ldots,p_n$ to desired $\ell_1$ distance $\epsilon,$ using a number of traces that scales polynomially with $n$ and $1/\epsilon,$ and succeeds with high probability in the random case where each $p_i$ is drawn independently from the uniform distribution over the interval $[0,1]$. 

Our algorithm is based on a scheme for identifying $w \approx log(n)$ sized ``chunks'' of traces that have \emph{no} deletions within them.  Crucially, we identify such chunks in a manner that does not look at the values of the trace within these regions---we identify such chunks only by looking at the values of the trace \emph{outside} the chunk in question.  Hence, the values within these deletion-free chunks are unbiased estimates of the true probabilities underlying these chunks.  Given this, the final straightforward step is to align these chunks---figure out the true indices of $S$ that gave rise to each of these chunks---then for each index $i\in \{1,\ldots,n\}$, return the average of the associated entries of the chunks. 

\vspace{-.3cm}
\begin{algorithm}[H]
\textbf{Algorithm 1: Find Deletion-Free Chunks:}\\
Input: Two sets of traces, $X$ and $Y$, with $|X|=m^2$ and $|Y| = m^{25}$, for some parameter $m>n$ with $m=\poly(n,1/\epsilon)$. \\
Output: Set of length $w = 10000 \log m$ ``chunks'', each of which is a contiguous region of a trace in set $X$. We will show that each such returned chunk corresponds to a deletion-free region. The set of traces $Y$ will be used solely to identify these chunks of traces in $X$.
\vspace{-2mm}
\begin{itemize}  
    \item For each trace $x \in X$, draw $i \in \{1,\ldots,n-3w+1\}$ uniformly at random, and consider the three consecutive length $w$ chunks beginning at index $i$: $L_x=x_{i,\ldots,i+w-1}$, $M_x=x_{i+w,\ldots,i+2w-1}$ and $R_x=x_{i+2w,\ldots,i+3w-1}.$ 
    \vspace{-2mm}
    \begin{itemize}
        \item We say that two length $w$ segments ``match'' if their $\ell_1$ distance is at most $5w/12$.  For each trace $y \in Y,$ and each index $j \in \{1,\ldots,n\}$, check whether $L_x$ matches $y_{j,\ldots,j+w-1}$, and similarly for $R_x$.
        \item Return the middle chunk, $M_x$, if the following two conditions hold:
        \begin{enumerate}
            \item There exists at least one $y \in Y$ and index $j$ such that $L_x$ matches $y_{j,\ldots,j+w-1}$ and $R_x$ matches $y_{j+2w,\ldots,j+3w-1}$.
            \item There do \emph{not} exist any $y \in Y$ for which $L_x$ matches $y_{j,\ldots,j+w-1}$ and $R_x$ matches $y_{j',\ldots,j'+w-1}$ with $j'-j > 2w$.  Namely, the only $y\in Y$ for which both $L_x$ and $R_x$ have matches must have the property that the locations they match to are offset by $\leq 2w$ indices.
        \end{enumerate}
    \end{itemize}
\end{itemize}
\end{algorithm}

\begin{algorithm}[H]
    \textbf{Algorithm 2: Align and Average Deletion-Free Chunks:}\\
    Input: A list of length $w$ binary sequences (corresponding to the output of the algorithm \emph{Find Deletion-Free Chunks}).\\
    Output: A length $n-2w$ vector of probabilities, $\hat{p}_{w+1},\ldots,\hat{p}_{n-w}.$  [The fact that we do not estimate the first and last $w$ probabilities is not an issue, as we can simply run these algorithms on traces that correspond to an instance that we have padded on each end by length $w$.]
    \begin{itemize}
    \item For a pair of length $w$ chunks in the input, $M,M'$, say that they ``match'' if their $\ell_1$ distance is at most $5w/12$, and say that they ``match with offset $1$''  if the $\ell_1$ distance between the last $w-1$ coordinates of $M$ and the first $w-1$ coordinates of $M'$ is at most $5w/12.$
    \item For all pairs of input chunks, check if they match or if they match with offset 1.  The matching relation partitions the input chunks into ``groups'', with each group defined as the chunks that match a given chunk.  Order these groups such that consecutive groups match with offset $1$.  (If this does not yield a total ordering, then return FAIL.)
    \item We now claim that for each chunk in the $i$th group, the bit at location $j$ is close to an unbiased estimate for the $p_{i+j-1}$.  For each $k$, our returned estimate $\hat{p_k}$ is simply the average of the coordinates of the chunks corresponding to estimates of $p_k$.
    \end{itemize}
\end{algorithm}

We will now analyze these algorithms.  We will end up proving that with $|X|=m^2$ and $|Y|=m^{25}$ as specified in Algorithm 1, with probability at least $1-1/\poly(m)$ over the randomness of the traces and true string $S$, we will recover $S$ to $\ell_1$ error $1/\poly(m)$.  (Note that we assume $m>n$.)  Given these fixed polynomials, $m$ can be set to the appropriate function of the desired error, $\epsilon$.  This parameterization in terms of $m$ simplifies the exposition and analysis.

We first argue that, with high probability, 1) for any two length $w$ chunks of traces that correspond to flips of the \emph{same} set of $w$ probabilities---namely that are perfectly ``aligned''---their $\ell_1$ distance will be at most $5w/12$ and hence they ``match'' and 2) If two chunks $z_{i,\ldots,i+w-1}$ and $z'_{j,\ldots,j+w-1}$ that are compared have $\ell_1$ distance at most $5w/12$ then there must at least one index that is aligned, in the sense that for some $k$, $z_{i+k}$ and $z'_{j+k}$ corresponded to flips of the same probability $p_{\ell}$ for some index $\ell$.  This will follow from a union bound over Hoeffding bounds.

\begin{lemma}\label{lem:hoeffding}
For a randomly generated $S=p_1,\ldots,p_n$, and two traces $x=x_1,x_2,\ldots$ and $y=y_1,y_2,\ldots,$ for any two chunks of length $w$  $x_{i,\ldots,i+w-1}$ and $y_{j,\ldots,j+w-1}$ let $q\le w$ denote the number of indices $t$ such that $x_{i+t}$ and $y_{j+t}$ originate from coin flips of the same probability $p_k$, and let $r=w-q$ denote the number of ``misaligned'' indices, namely where $x_{i+t}$ originated from $p_k$ and $y_{j+t}$ originated from $p_{k'}$ for $k \neq k'$.  Then the following concentration bound holds, where the probability is over the randomness of $S$ and the coin flips in the two traces (but not over the randomness of which bits are deleted, which determine $q,r$) $$\Pr\left[\left| \sum_{t=0}^{w-1} |x_{i+t}-y_{j+t}| - \left(\frac{q}{3}+\frac{r}{2}\right) \right| \geq \beta\right] \le 2 \exp\left(\frac{-2\beta^2}{w}\right).$$
\end{lemma}
\begin{proof}
    We step through the terms $|x_{i+t}-y_{j+t}|$ one by one, in order of increasing $t$.  If both bits correspond to flips of the same probability, $p_k$, then no earlier terms in this sum could have corresponded to $p_k$, and hence the contribution of this sum is independent of the contributions of the previous terms. 
    The expected value (with respect to the randomness of drawing $p_k$ and the realization of these flips) is $\int_{0}^1 2 p(1-p) \d p = 1/3$. 

    In the case that the two bits $x_{i+t}$ and $y_{j+t}$ correspond to realizations of different probabilities in the true string, $p_k,p_{k'}$ for $k \neq k'$, then note that at least one of these probabilities must not have been encountered in the sum thus far.  Without loss of generality assume that is $p_k$, which is the probability corresponding to $x_{i+t}$.  Whatever the value of $y_{j+1}$, the probability $x_{i+t}=y_{j+t}$ is trivially $1/2$, with respect to the randomness of $p_k$ and the realization of $x_{i+k}$, and is independent of the previous terms in the sum, as $p_k$ is drawn independently of the probabilities encountered previously in the summation. 
    
    Hence the quantity in question, $\|x_{i,\ldots,i+w-1}-y_{j,\ldots,j+w-1}\|_1$ corresponds to a sum of independent $0/1$ random variables and has expectation $\frac{q}{3}+\frac{r}{2}$.  The claimed concentration now follows from the standard Hoeffding bound applied to sums of $w$ independent 0/1 random variables. 
\end{proof}

We now take a union bound over the above lemma to argue that, with high probability, all chunks that are perfectly aligned will match,  and all matches have at least one aligned index:
\begin{lemma}\label{lemma:match}
The following holds with probability at least $1 - (2|X||Y|n + 3|X|^2)e^{-w/72} \ge 1-O(1/m^{100})$: For each of the $\le 2|X|\cdot|Y|n$ pairs of chunks whose $\ell_1$ distance is computed in the ``Find Deletion-Free-Chunks'' algorithm, and each of the $\le 3|X|^2$ chunks whose distance is computed in the ``Align and Average Deletion-Free Chunks'' algorithm, if the pair of chunks have \emph{no} aligned indices then their $\ell_1$ distance will be greater than $5w/12$, and if the pair of chunks have perfect alignment, then their $\ell_1$ distance will be less than $5w/12$.
\end{lemma}
\begin{proof}
This is a union bound over the previous lemma with $\beta = w/12$, with the union bound accounting for the total number of times chunks are compared in the two algorithms.  The factor of 2 is because each $x\in X$ has two chunks that get compared, namely $L_x$ and $R_x$, and the factor of $3|X|^2$ accounts for the fact that in this second algorithm we compare chunks with a possible offset of $-1,0,$ and $1$.  
\end{proof}

Throughout the rest of the proof, we will assume that stipulation in the lemma holds, namely that all matched chunks do have at least one aligned index, and all perfectly aligned chunks that are compared will match.

We now argue that, with high probability, the first algorithm returns at least $\poly(m)$  deletion-free chunks corresponding to each of the $\le n$ possible windows of length $w$ of the original string S=$p_1,\ldots,p_n$ (excluding windows that overlap the first or last $w$ indices).  The next lemma asserts that for each $i\in \{1,\ldots,n-3w+1\}$ there will be $\poly(m)$ $x \in X$ for which $L_x, M_x$ and $R_x$ correspond to a contiguous deletion-free block of $S$ beginning at index $i$, namely correspond to the probability $p_i,\ldots,p_{i+3w-1}$.  The given that, the subsequent lemma argues that, with high probability, \emph{all} such middle chunks $M_x$ will be returned by that algorithm.

\begin{lemma}\label{lemma:lem6}  
With probability $1-\exp(-\Omega(m^{0.9}))$, for each index $i \in \{1, n - 2w\}$, there are at least $\poly(m)$ deletion-free chunks $L_x,M_x,R_x$ corresponding to probabilities $p_i, \ldots, p_{i + 3w - 1}$.
\end{lemma}
\begin{proof}
The expected number of traces in $X$ for which $L_x,M_x,R_x$ have no deletions and correspond to the desired probabilities is $\ge |X|\frac{1}{n}(1-\del)^{3w}= \Omega(m^{0.9})$.  Hence by a Chernoff bound, the probability that there are fewer than half this expected number is inverse exponential in $m^{0.9}$.
\end{proof}

\begin{lemma}\label{lemma:lem7}
Given an $x\in X$ for which $L_x, M_x$ and $R_x$ correspond to a contiguous deletion-free block of $S$ beginning at index $i$, namely those chunks correspond to the probabilities $p_i,\ldots,p_{i+3w-1}$, the probability the first algorithm fails to return $M_x$ is at most inverse exponential in $m$.
\end{lemma}
\begin{proof}
First,  we will show that, with high probability there will be a $y\in Y$ that also has a deletion-free region corresponding to $p_i,\ldots,p_{i+3w-1}$, and hence $L_x$ and $R_x$ will match the associated regions of $y$ (by the assertion after Lemma~\ref{lemma:match}), and since there are no deletions between these regions in $x$ or $y$, these regions in $y$ will be separated by width exactly $w$, satisfying the condition 1) in the algorithm.
As in the previous lemma, the expected number of traces in $Y$ with no deletions in this region is $|Y|(1-\del)^{3w} \gg m$, and hence by a Chernoff bound, the probability no such $y\in Y$ has this property is inverse exponential in $m$.

Next, we show that the second condition of the algorithm is satisfied, namely that $L_x$ and $R_x$ will not match any regions of $y$ that are separated by more than $w$ indices. Consider a $y$ that has a chunk matching $L_x$ and a chunk matching $R_x$. Let $L_y$ denote the chunk of $y$ that matches $L_x$, let $R_y$ denote the chunk of $y$ that matches $L_x$, and let $M_y$ be the bits of $y$ in between $L_y$ and $R_y$. Note that by the assertion after Lemma~\ref{lemma:match}, at least one index of $L_x$ must be aligned with the corresponding index of $L_y$, and similarly at least one index of $R_x$ must be aligned with the corresponding index of $R_y$. Because the relevant chunks of $x$ are deletion free, there cannot be more bits in between these two locations in $y$ than in $x$. Therefore the middle region $M_y$ must have at most $w$ bits. Hence, $L_x$ and $R_x$ cannot match any regions of $y$ that are separated by more than $w$ bits.
\end{proof}

Thus far, we have proved that the first algorithm will return at least $\poly(m)$ deletion-free chunks, $M_x$, for each offset.  What remains is to prove that with high probability, every chunk returned will be deletion-free.  To this end, we now show that with high probability, every chunk $L_x$ and $R_x$ considered has at most $d=10\frac{\log m}{\log 1/\del}$ deletions.  Furthermore, we show that if $M_x$ had any deletions, there will be some $y\in Y$ that satisfies the following 1) $y$ has the same pattern of deletions in the regions associated with $L_x$ and $R_x$ and hence will match with their respective regions $L_y$ and $R_y$ and 2) $L_y$ and $R_y$ are separated by at least $w+1$ containing a $w+1$ sized deletion-free subset of the $w$ probabilities that contributed to $M_x$ and the $\ge 1$ deleted bits/probabilities within $M_x$.

\begin{lemma}\label{lem:8}
With probability at least $1-O(m^{-0.2})$, no trace in $X$ has more than $d=10\frac{\log m}{\log 1/\del}$ deletions within either $L_x$ or $R_x$.

\end{lemma}
\begin{proof}
Given a number of deletions $j\geq d$, the probability that a fixed substring of length $w+j$ ends up with exactly $j$ deletions so that it results in exactly $w$ bits of a trace equals ${w+j\choose j} \del^j (1-\del)^w$. The probability that this ever happens, over all $n$ locations in the string, and over all $|X|$ traces, is at most \[n |X|\sum_{j=d}^\infty {w+j\choose j} \del^j (1-\del)^w\]
The ratio of consecutive terms of the sum is $\del\frac{w+j}{j}\leq \del\frac{w+d}{d}$, which is at most $\frac{1}{2}$ since $d\geq \frac{w}{\frac{1}{2\del}-1}$. In this case the total probability is bounded by \[2n |X| {w+d\choose d} \del^d (1-\del)^w.\]

We simplify the ${w+d \choose d}$ term  via Stirling's approximation, which, up to lower order terms, yields the following expression:
$2n|X|\left(\frac{w}{d}\right)^{d} (1+d/w)^{w+d} \del^d (1-\del)^w$.
Plugging in the prescribed values of $|X|,w,$ and $d$, and setting $\del=10^{-7}$ as the bound is monotonically decreasing as $\del$ decreases, yields that this expression is at most $2m \cdot m^2 \cdot m^{6.1} \cdot m^{0.7} \cdot m^{-10}=O(m^{-0.2}).$ 
\end{proof}

\begin{lemma}\label{lemma:onlyDF}
With probability at least $1-O(m^{-0.2})$, all $M_x$ returned by the first algorithm will be deletion-free. 
\end{lemma}
\begin{proof}
By the previous lemma, with at least the claimed probability,  for all $x \in X$, $L_x$ and $R_x$ each contain $\leq d$ deletions. Henceforth, we assume this holds. 

Consider a trace $x$ where $M_x$ has at least one deletion. Let $d_1,d_2,d_3$ denote the number of deletions in $L_x,M_x,R_x$ respectively, and note that by assumption $d_1,d_3 \le d$ and $d_2 \ge 1.$  We now consider the probability that a trace contains 1) the \emph{exact} same pattern of $d_1$ deletions within the region corresponding to $L_x$ and the same pattern of $d_3$ deletions within the region corresponding to $R_x$, and 2) contains no deletions among the first $w+1$ probabilities in the range between the probability $p_k$ corresponding to the first coordinate of $M_x$ and the probability $p_{k'}$ corresponding to the last coordinate of $M_x$.  Since there is a deletion in $M_x,$ this range must contain at least $w+1$ elements.   If at least one such trace is in $Y$, by the assertion after Lemma~\ref{lemma:match}, $L_x$ and $R_x$ will match the corresponding regions of that trace, yet their separation will be at least $w+1$, and hence $M_x$ will not pass the second condition of the algorithm and will not be returned. 

The probability that no such $y\in Y$ occurs is $$\le \left(1- (1-\del)^{w + w + (w+1)} \del^{d_1+d_3}\right)^{|Y|} \le \left(1- (1-\del)^{3w+1} \del^{2d}\right)^{|Y|}\le \exp\left(-|Y| (1-\del)^{3w+1} \del^{2d} \right).$$
Plugging in the prescribed expressions for $|Y|$ and $d$ and the bound that $\del < 10^{-7}$ yields that this probability is: $$\exp(-|Y| (1-\del)^{3w+1} \del^{2d})=\exp(-m^{25} (1-\del)^{30000\log  m + 1} \del^{-20\log m /\log \del}) \le \exp(-m^{25}\cdot (1/2) \cdot m^{-20})=\exp(-\Omega(m^5)).$$
\end{proof}

To complete our proof of Theorem~\ref{thm:upper_bound}, we now consider the second algorithm, ``Align and Average Deletion-Free Chunks''.  Given that Algorithm 1 only returns deletion-free chunks (Lemma~\ref{lemma:onlyDF}), two such chunks are either completely aligned, or have zero aligned indices.  Hence, by Lemma~\ref{lemma:match}, for any chunks that align (with either 0,1 or $-1$ offset), they do originate from identical regions of $S$ (or offset by 1 or $-1$, respectively).  Additionally, the first algorithm never looks at the contents of the returned chunks (it just looks at the adjacent chunks $L_x,R_x$), hence the values in the returned chunks are independent unbiased estimates of their underlying probabilities.  If we knew which indices of $S$ corresponded to each of these chunks, we would have independent unbiased estimates of each $p_k$.  Algorithm 2 figures out this correspondence, and the independence no longer holds after conditioning on the successful alignment of Algorithm 2.  However, we have shown that this successful alignment happens with high probability, and hence the coordinates of the chunks are close to independent unbiased estimates of these probabilities (and in particular, with high probability, cannot be distinguished from independent realizations).

The claim that Algorithm 2 successfully aligns these chunks with high probability follows from Lemmas~\ref{lemma:lem6} and~\ref{lemma:lem7}.  In that case, the alignment will be successful and each index will be estimated to accuracy $1/\poly(m)$.  Taking $m$ to be a sufficiently large polynomial of the reciprocal of the desired accuracy and $n$, completes the theorem, modulo the question of recovering the first and last $w$ indices.  To accomplish this, one could simply ``pad'' each trace, such that the padded trace corresponds to an instance of the generalized trace reconstruction problem of size $n' = n+2w,$ whose middle $n$ coordinates correspond to the true string $S$.

\paragraph{Acknowledgments}
GV is supported by NSF Award AF-2341890 and a Simons Foundation Investigator Award.  GV is currently affiliated with OpenaAI but this work was done at Stanford.
PV is partially supported by NSF award CCF-2127806 and by Office of Naval Research award N000142412695.
\bibliographystyle{alpha}
\bibliography{refs}
\end{document}